\documentclass[twocolumn]{aastex631}
\usepackage[linesnumbered,lined,boxed,commentsnumbered]{algorithm2e}

\begin{document}

\title{Automatic Parallel Tempering Markov Chain Monte Carlo with Nii-C}

\correspondingauthor{Sheng Jin}
\email{jins@ahnu.edu.cn}

\author[0000-0002-9063-5987]{Sheng Jin}
\affiliation{Department of Physics, Anhui Normal University, Wuhu, Anhui 241002, China}
\affiliation{Purple Mountain Observatory, Chinese Academy of Sciences, Nanjing, Jiangsu 210023, China}

\author{Wenxin Jiang}
\affiliation{Department of Statistics and Data Science, Northwestern University, Evanston, IL 60208, USA}

\author[0000-0001-9424-3721]{Dong-Hong Wu}
\affiliation{Department of Physics, Anhui Normal University, Wuhu, Anhui 241002, China}

\begin{abstract}
Due to the high dimensionality or multimodality that is common in modern astronomy, sampling Bayesian posteriors can be challenging. Several publicly available codes based on different sampling algorithms can solve these complex models, but the execution of the code is not always efficient or fast enough.
The article introduces a C language general-purpose code, Nii-C\footnote{\url{https://github.com/shengjin/nii-c.git}} \citep{Jin2024}, that implements a framework of Automatic Parallel Tempering Markov Chain Monte Carlo.
Automatic in this context means that the parameters that ensure an efficient parallel tempering process can be set by a control system during the inital stages of a sampling process.
The auto-tuned parameters consist of two parts, the temperature ladders of all parallel tempering Markov chains and the proposal distributions for all model parameters across all parallel tempering chains.
In order to reduce dependencies in the compilation process and increase the code's execution speed, Nii-C code is constructed entirely in the C language and parallelised using the Message-Passing Interface protocol to optimise the efficiency of parallel sampling.
These implementations facilitate rapid convergence in the sampling of high-dimensional and multi-modal distributions, as well as expeditious code execution time.
The Nii-C code can be used in various research areas to trace complex distributions due to its high sampling efficiency and quick execution speed.
This article presents a few applications of the Nii-C code.
\end{abstract}


\section{Introduction} \label{sec:intro}

Bayesian inference is a powerful tool for fitting or evaluating user-defined models to a set of data.
It provides a statistical approach to estimating model parameters by summarizing the results in terms of probability distributions.
Thanks to the recent advances in numerical methods and computing power, Bayesian inference can be easily implemented numerically and is therefore widely used in almost all scientific disciplines.

In modern astronomy, Bayesian inference has particularly elegant applications.
It is the standard framework in the field of gravitational wave astronomy for inferring the astrophysical properties of the sources \citep{Thrane2019,Ashton2019,Smith2020,Breschi2021}; 
it is a self-consistent statistical way to constrain complex models that combine heterogeneous observations in exoplanet science 
\citep{Gregory2005a,Gregory2005b,Ford2006,Spiegel2012,Diaz2014,Rogers2015,Wolfgang2016,Dorn2017,Parviainen2018,Sestovic2018,Feng2019,Li2021,Jin2021,Huang2023};
it provides a data-driven approach to extracting intrinsic stellar properties from observed spectra
\citep{Pont2004,Jorgensen2005,Gruberbauer2009,Burneet2010,Kallinger2010,Handberg2011,Sale2012,Binney2014,Ness2015};
it is widely used for model selection and parameter estimation in 
the study of the Milky Way and cosmology 
\citep{Mukherjee2006,Schonrich2010,Bovy2012,Rix2013,Piffl2014,Kafle2014,Bland2016,McMillan2017}, etc.

The main objective of Bayesian inference is to find the posterior probability distribution, and Markov Chain Monte Carlo (MCMC) is an effective method for achieving this.
Publicly available MCMC codes make it easy to get started with Bayesian analysis of research topics in astronomy and astrophysics \citep{Sharma2017,Parviainen2018}.
An important category of these tools is general-purpose codes that implement one or more sampling algorithms to provide the users with a calling interface, for example, PyMC3 \citep{Salvatier2015}, STAN\citep{Stan2023}, EDWARD \citep{Tran2016}, emcee \citep{Foreman2013}, ZUES \citep{Karamanis2021}, and so on.
These packages offer the scientific community modularized, well-structured, and ready-to-use Bayesian analysis tools.
However, the high-dimensional and multimodal models in certain astronomical models present convergence challenges for these general-purpose codes.
The first reason is that a sampling process tends to get stuck near local density extremes in complex distributions, and the second is that the difficulty of finding suitable proposal distributions 
for all model parameters increases with the dimensionality of a parameter space.
In order to overcome these difficulties, a variety of sampling algorithms have been developed for various astronomical applications, like the Nested sampling methods that directly give Bayesian evidence \citep{Skilling2004,Skilling2006,Mukherjee2006,Parkinson2006,Feroz2009,Brewer2011,Speagle2020,Buchner2023}.

As a concise and efficient sampling algorithm, MCMC also incorporates a number of techniques to facilitate convergence for complex distributions.
Parallel tempering algorithms provides a concise approach for escaping local maximum regions \citep{Liu2001}. 
When combined with an automatic or adaptive control system for adjusting the sampling proposals, such a set of algorithms can achieve fast convergence for complex models \citep{Gregory2005b,Miasojedow2013}.
These algorithms are referred to as Automatic (or Adaptive depending on the implemention of its tuning method) Parallel Tempering Markov chain Monte Carlo (APT-MCMC).
Publicly available APT-MCMC codes are provided in the programming language of Mathematica \citep{Gregory2005a}, Matlab \citep{Miasojedow2013}, Julia \citep{Vihola2020} and Python \citep{Ellis2019,Vousden2021}.
APT-MCMC code developed in the C language  is necessary to handle larger amounts of data at higher execution speeds and to facilitate integration with various existing codes for different physical models.

Here, Nii-C is introduced as a C language version general-purpose parallel tempering MCMC code. 
It has an automatic control system that can dynamically adjust the temperature ladders of all parallel tempering Markov chains and the proposal distributions of all model parameters of all the chains in an early tuning stage. 
Such a set of algorithms is referred to in this article as Automatic Parallel Tempering Markov chain Monte Carlo.
The Nii-C code is implemented in pure C programming language using the Message-Passing Interface (MPI) parallelization to reduce the dependencies in the compilation of the code and to improve its runtime speed.
In order to provide a general-purpose framework for analyzing and interpreting data in a variety  of research areas, Nii-C puts the prior and likelihood functions into separate source files to facilitate the implementation of different posterior distributions.
This article demonstrates the convergence capability of these algorithms through a number of applications and benchmark comparisons.

This article is organized as follows.
Section \ref{sec:algo} describes in detail the parallel tempering algorithms and the initial control system that aims to automatically adjust 
the temperature ladders and the proposal distributions for all model parameters of all parallel chains.
Section \ref{sec:eg} presents a number of example applications and benchmark comparisions using the Nii-C code, a simple liner regression model, a hierarchical Bayesian model to infer the radius relationship of hot Jupiters, a planetary orbital parameter fitting model, a multimodal eggbox distribution, a multimodal curved likelihood, and a Rastrigin function distribution.
A brief discussion is given in Section \ref{sec:diss}.

\section{Algorithms}

\label{sec:algo}
When sampling high-dimensional or multimodal distributions, there are two main difficulties that must be solved to achieve convergence.
First, samplers tend to get stuck at local extremes and thus are unable to reach the global extreme. 
Second, it is not possible to set the proposal distributions for all model parameters to guarantee both an efficient searching of the global parameter space early on and a suitable sampling acceptance rate later on.
The Nii-C code has implemented two specific techniques to solve these two problems. These are the parallel tempering algorithm to help the samplers jump out of local extremes, and a control system in an early tuning stage to 
automatically adjust the parallel temperature ladders and the proposal distributions for all model parameters of all parallel chains.

\subsection{Parallel Tempering MCMC}  \label{sec:ptmcmc}

In most Bayesian applications, a target posterior probability distribution is sampled in order to infer the values of the model parameters or to evaluate the correctness of different models, which is of the form:

\begin{equation}
p(\theta|D,M) \propto p(D|\theta,M)\,p(\theta|M)\,
\label{eqn:bayes2}
\end{equation}

where $\theta$ is the set of parameters need to be estimated with a particular model $M$ for the observed data $D$, $p(D|\theta,M)$ is the likelihood reflecting the probability of the observed data $D$ occurring if the parameters in model $M$ are equal to $\theta$, and
$p(\theta|M)$ is the prior distribution of $\theta$, representing the beliefs about the parameters before observing the data.

Although posterior distributions are easy to construct in Bayesian applications,  they do not always converge with an MCMC sampler \citep{Tierney1994,Mengersen1996}.
This is particularly true when sampling a high-dimensional distribution, where an MCMC sampler can easily get stuck in local optimal solutions and not fully explore the global distribution.

Parallel tempering, the technique used in Nii-C, is introduced as an effective method to overcome the trapping in local extremes \citep{Geyer1995,Liu2001,Gregory2005b}.
In this method, a set of flatter versions of the posterior distributions of interest are generated using a parameter $\beta$:

\begin{equation}
        \widehat{p}(\theta|D,\beta,M) \propto p(D|\theta,M)^{\beta}\,p(\theta|M)\,, \quad {\mathrm {for}} \,\, \, 0<\beta<1
\label{eqn:temp}
\end{equation}

where $\beta$ equal to 1 corresponds to the original posterior distribution, and $\beta$ equal to 0 corresponds to the posterior distribution consisting of a completely flat likelihood.
In this way, local optimal solutions can be easily escaped by an MCMC sampler with an appropriate $\beta$ value.

At random sampling states, let these states be denoted by $s$, a pair of adjacent Markov chains with parameters $\beta_i$ and $\beta_{i+1}$ are randomly selected and their current positions in the parameter space $\theta_{s,i}$ and $\theta_{s,i+1}$ are swapped with a probability of 
\begin{equation}
        r = {\mathrm {min}} \Biggl[\,1, \, \frac{ \widehat{p}\big(\theta_{s,i+1}|D,\beta_i,M\big) \, \widehat{p}\big(\theta_{s,i}|D,\beta_{i+1},M\big)}{ \widehat{p}\big(\theta_{s,i}|D,\beta_i,M\big) \, \widehat{p}\big(\theta_{s,i+1}|D,\beta_{i+1},M\big)} \, \Biggr]\,
\label{eqn:pltemp}
\end{equation}

Such a swap of the positions of the parallel samplers can effectively speed up the global search of the entire parameter space.

\subsection{Early Tuning Stage of Temperature Ladders} \label{sec:automatic1}

The efficacy of the parallel tempering MCMC algorithm is contingent upon the selection of an optimal combination of $\beta$ values.
Traditionally, a temperature parameter is introduced as $T = 1/\beta$.
In this context, an efficient parallel tempering MCMC requires chains with sufficiently high temperatures to escape local extremes and an appropriate temperature ladder to assist the sampler of the target distribution in exploring the entire parameter space.
Researches suggested that a favourable combination of temperatures will result in uniform acceptance ratios between different pairs of adjacent chains \citep{Sugita1999,Kofke2002,Earl2005}.
It can be expected that the most optimal combination of temperature ladders will vary according to the specific sampling model that is employed.
\citet{Vousden2016} have developed an selection algorithm that can dynamically adjust the number of parallel tempering chains and the chain temperatures in order to achieve uniform acceptance ratios between neighbouring chains.
This dynamic temperature selection improvement demonstrates superior performance compared to parallel tempering MCMC with fixed temperature configurations. 

The current version of Nii-C code does not permit the number of parallel tempering chains to be altered during a sampling process.
A straightforward approach to adjusting the temperature gradient was devised in the Nii-C code, namely maintaining the hottest and coldest temperatures constant while adjusting the temperature intervals of the intermediate parallel tempering chains. 
The objective of each adjustment is to modify the temperature of the middle chain of every three neighbouring chains, with the intention of reducing the difference in the swap acceptance ratios obtained by the parallel tempering processes between the middle chain and a chain on either side.
This is achieved by moving the temperature of the intermediate chain towards the side which has a lower swap acceptance ratio.

It should be noted that the swap acceptance ratios between parallel tempering chains depend not only on the temperature ladders, but also on the real-time positions of the parallel samplers. 
Therefore, even when constant temperature ladders are used, the swap acceptance ratios between parallel tempering chains will fluctuate dynamically over time.
With this in mind, the adjustment of the temperature ladders in a parallel tempeiring MCMC need not be a lengthy process and is often unnecessary for simple distributions.

Although adjusting the temperature ladders improves the efficiency of the parallel tempering MCMC algorithm, it violates the Markovian property because the state of the next iteration in a Markov chain can only depend on the current state \citep{Grimmett2001}.
Such a chain fails the principle of detailed balance and will not converge to a stationary distribution.
Therefore, Nii-C keep the adjustment of temperature ladders in an early tuning stage and provides an input parameters to set the end of this stage.
The configuration of the automatic temperature ladders selection process in the overall Nii-C workflow will be described in detail in session \ref{sec:workflow}.

\subsection{Early Tuning Stage of Proposal Distributions} \label{sec:automatic2}

To achieve fast convergence, the acceptance rate of an MCMC sampler should be appropriate to obtain enough effective sampling points when performing random walks. 
A low acceptance rate means that the sampler is rejecting the majority of proposals, leaving too few effective points for convergence.
Too high an acceptance rate means that the sampler is not going far enough and may not be exploring the parameter space efficiently.
For the Metropolis algorithm, the optimal acceptance rate for multi-parameter models is $\sim$ 23.4\%  \citep{Gelman1996,Gelman1997}.

The main determinant of a sampler's acceptance rate is the Metropolis-Hastings step sizes of all parameters, characterised by the standard deviation of the Gaussian proposal distribution when new proposals are made in sampling. 
For a successful MCMC, the proposal distributions for all model parameters should be large at the beginning of the random walk to help a sampler fully explore the entire parameter space.
Then, as the sampler enters the region where most of the posterior probability mass is located, these proposal distributions should decrease accordingly to allow the sampler to build up an accurate map of the areas that has the highest probability mass.

However, it is difficult to implement such an ideal dynamic process of Metropolis-Hastings proposal distributions in the random walk of an MCMC.
Because an appropriate combination of proposal distributions not only requires separate settings for each of the model parameters, but also depends on the instantaneous position in the parameter space being traversed by a sampler.
Consider a model with $M$ parameters whose posterior distribution is being sampled by parallel tempering MCMC with $N$ chains with different $\beta$ values.
In this case, to achieve a reasonable acceptance rate requires adjusting and monitoring $M \times N$ Metropolis-Hastings proposal distributions in real-time while the sampler randomly  changes its position in the target distribution. 
Manual handling of this process is not feasible.

The Nii-C code has implemented a control system that can automatically adjust all Metropolis-Hastings proposal distribtutions in the early tuning stage to ensure an optimal acceptance rate in sampling. 
Algorithm \ref{alg:one} describes the algorithm of this automatic tuning process.
Please note that the algorithm includes a setting (line 52 in Algorithm \ref{alg:one}) that can randomly revert the Gaussian proposals of all parameters to one-tenth of their respective prior ranges in specific cases.
Experiments have shown that this setting is crucial for ensuring quick convergence for complex distributions.

Similarly to the automatic adjustment of temperature ladders, scaling the sizes of proposal distributions also violates the Markovian property.
To correctly use a changing combination of proposal distributions, adaptive MCMC algorithms are specially designed in a careful way that asymptotically the change is zero, so that the chain is ergodic and converges to the target distribution \citep{Haario2001,Andrieu2008}.
An alternative method of satisfying the Markovian property of a chain is to disable the tuning stage once the estimated acceptance rate falls within the ideal range \citep{Gregory2005a,Gregory2005b}.
This allows the sampler to operate as the standard parallel tempering algorithm thereafter, when the swap moves between different temperatures satisfy detailed balance for the correct target distribution on a product space \citep{Woodard2009}.
In this case, geometric ergodicity is guaranteed on the basis of Theorem 1 in \citet{Miasojedow2013}. 
Nii-C restricts the scaling of proposal distributions in an early tuning stage, as detailed in the next subsection.

\RestyleAlgo{ruled} 
\begin{algorithm*}
\caption{Algorithm for tuning the proposal distributions of a model with $N_{para}$ parameters sampled by $N_{rank}$ parallel tempering chains.}\label{alg:one}
\KwData{The old sampling step size matrix $S$ ($N_{rank}$ $\times$ $N_{para}$)}
\KwResult{The updated matrix $S$}
 $IdealAcceptanceRate \gets 0.234$\;
 $AllowableError \gets$ ``\texttt{preset value}"\;
 run a stack of the parallel tempering MCMC\;
 \For{each rank $i$ in $N_{rank}$}{
   \If{the acceptance rate of the stack is not in the right range}{
     create a matrix $A$ ($N_{para}$ $\times$ $N_{resized}$) to store the acceptance rates of all tuning chains\;
     create a matrix $R$ ($N_{para}$ $\times$ $N_{resized}$) to store the resized sampling step sizes\;
     \For{each model parameter $j$ in $N_{para}$}{
        $R[j][:] \gets$ a group of resized $S[i][j]$\;
        \For{each $k$ in $N_{resized}$}{
         create a temporary step size array $T$ $(N_{para})$\;
         $T[:] \gets S[i][:]$ \;
         $T[j] \gets R[j][k]$\;
         run a test chain using $T$ at the parameter space where the $rank$ $i$ is sampling with its own $\beta$\;
         $A[j][k] \gets$ the acceptance rate of the test chain\;  
       }  
     }
     create an array $D$ $(N_{para})$\;
     \For{j in $N_{para}$}{
     $D[j] \gets$ the standard deviation of $A[j][:]$\;
     }
     $a \gets$ argmax($D$)\;
     $b \gets$ argmin($A[a][:]-IdealAcceptanceRate$)\;
     $c,d \gets$ argmin($A[:][:] - IdealAcceptanceRate$)\;
     \eIf{$min(A[a][:]-IdealAcceptanceRate)$ $<$ $AllowableError$}{
       \eIf{$min(A[:][:]-IdealAcceptanceRate)$ $<$ $AllowableError$}{
         $u \gets U(0,1)$\;
         \eIf{$u > 0.5$}{
            $J,K \gets a,b$\;
         }{
            $J,K \gets c,d$\;
         }
       }{
          $J,K \gets a,b$\;
       }  
     }{
       \eIf{$min(A[:][:]-IdealAcceptanceRate)$ $<$ $AllowableError$}{
         $J,K \gets c,d$\;
       }{
         $u \gets U(0,1)$\;
         \eIf{$u > 0.5$}{
           $J,K \gets -1,0$\;
         }{
           $J,K \gets -1,-1$\;
         }
       }
     }
   \eIf{$J \geq 0$}{
     $S[i][J] \gets R[J][K]$\;
   }{
     \eIf{$K \geq 0$}{
       reset $S[i][:]$ to one-tenth of the corresponding prior range\;
     }{
       \For{j in $N_{para}$}{
         $S[i][j] \gets R[j][$argmin$(A[j][:] - IdealAcceptanceRate)]$\;
       }
     }
   }
   }
 }
\end{algorithm*}

\subsection{The Nii-C workflow}
\label{sec:workflow}

The Nii-C code splits an entire Markov chain into two parts, following the approach used by Gregory \citep{Gregory2005a,Gregory2005b}.
Initially, there is an early tuning stage where a control system automatically adjusts 
the temperature ladders of all parallel tempering Markov chains and the proposal distributions for all model parameters across all parallel tempering chains.
The tuning stage can ensure an appropriate combination of temperature ladders and
induce a good acceptance rate when the sampler explores the target parameter space.
Subsequently, the control system turns off the initial non-Markovian tuning stage and returns to the standard parallel tempering algorithm.
This allows the sampler to meet the detailed balance requirements and garantees that the chains converge to stationary distributions.
As parallel tempering continues in the later phase, the sampler can still effectively explore the global parameter space with fixed proposal distributions. 
Additionally, by repeating the whole procedure for a few different choices of initial parameters and comparing the trace plots of each parameter in different runs, the convergence of the chains can be verified.

\begin{figure*}[ht!]
\centering
   \includegraphics[width=5.00in]{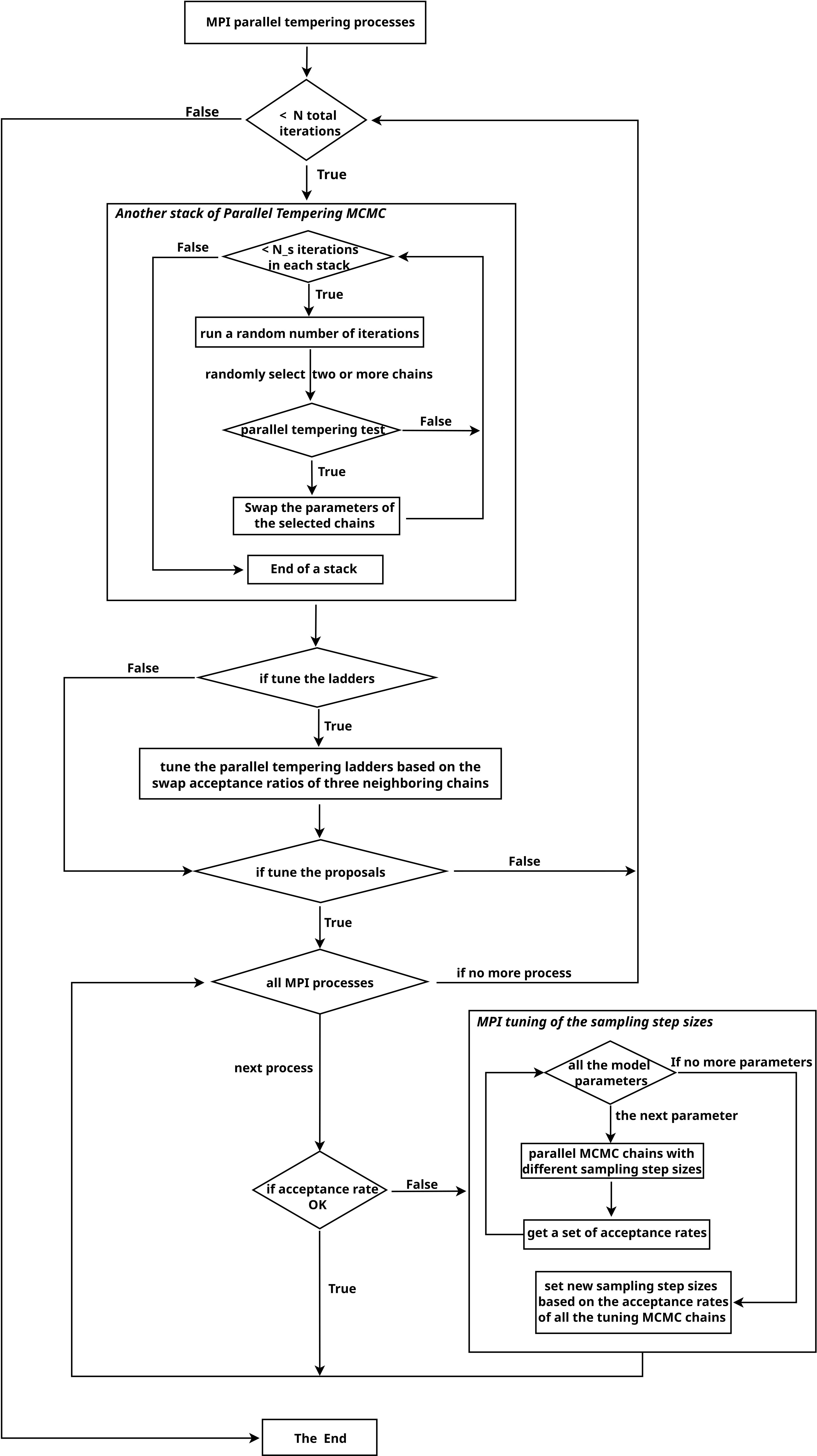}
\caption{The workflow of Nii-C code.}
\label{fig:one}
\end{figure*}

Figure \ref{fig:one} shows the workflow of the Nii-C code.
At the start of a sampling process, Nii-C sets up a group of $N_{rank}$ parallel Markov chains with different values of $\beta$, representing different tempered states of a target posterior distribution.
The total number of iterations in an APT-MCMC process is split into many stacks of short iterations.
In each stack, multiple parallel tempering chains run simultaneously and their iterations are further segmented into many short sequences of random length, which are called batches.
At the end of each batch, two or more chains are selected randomly and tested for swapping using Equation \ref{eqn:pltemp}.

In the Nii-C code, the initial stages of automatic tuning of temperature ladders and proposal distributions are designed as optional processes that can be controlled by initial input parameters.
If the option to adjust the temperature ladders is switched on, the swap acceptance ratios of neighbouring parallel tempering chains are calculated after each stack of iteration in an initial tuning stage.
Based on the calculated swap acceptance ratios between neighbouring parallel tempering chains, the temperature ladders will be automatically adjusted according to the method described in section \ref{sec:automatic1}.
    
If the option to adjust the proposal distributions is switched on,
Nii-C computes the sampling acceptance rates of all the parallel tempering chains after each stack of iterations in the initial tuning stage.
It identifies chains with a bad acceptance rate, either too high or too low, as target chains that require tuning of the sampling  proposals. 
To tune each of the target chains, Nii-C carries out many independent short Markov chains with different combinations of sampling proposals for all model parameters. 
Then, Nii-C calculates the acceptance rates of all these test Markov chains.
Based on these acceptance rates, Nii-C sets a new combination of proposals using Algorithm \ref{alg:one}.
Typically, stacks of iterations and tuning processes are performed alternately at the beginning of the tuning stage.
Subsequently, when the samplers of all the parallel tempering chains achieve good acceptance rates around convergence regions, the combinations of the proposal distributions begin to stabilize.
At this stage, further tuning is unnecessary.

To ensure that the Markovian property can be preserved in parallel tempering chains, Nii-C sets a limit on the number of iterations after which the early tuning stage of proposal distributions is switched off.
In fact, even if this limit number is not explicitly declared, the tuning stage will be stopped quickly as the sampler achieve good acceptance rates around convergence regions.
Models with a moderate number of parameters typically require only a few tuning cycles at the start of the sampling process.
However, Nii-C requires setting a limit on the number of iterations to avoid any unexpected tuning cycles that could later compromise the Markovian property.

Note that care should be taken to avoid adjusting the temperature ladders at the same time as tuning the proposal distributions of model parameters.
Because a suitable combination of proposal distributions is associated with a particular temperature value. 
Adjusting both at the same time can lead to chaotic results, as a newly set temperature value will invalidate a previously well-tuned combination of proposal distributions.
In general, the temperature ladders should be adjusted at the earliest stage with a fixed combination of proposal distributions, then the temperature adjusting should be switched off to conduct automatic proposal distribution tuning, and finally proposal tuning should also be switched off for the bulk process of parallel tempering MCMC.

In essence, the Nii-C code aims to be a flexible parallel MCMC code that can be conveniently adapted to diverse requirements by a range of users.
If we set the initial $\beta$ values to 1 for all parallel Markov chains and disable the tuning of temperature ladders and the proposal distributions, then Nii-C transformes into a basic parallel MCMC code featured with a fast execution speed. 
If we turn off the automatic tuning of temperature ladders, as in cases where an empirical combination of temperature values is known for a target distribution, Nii-C becomes a conventional parallel tempering MCMC code.
Additionally, Nii-C provides special settings that allow the execution process of a parallel MCMC to be controlled in more detail.
For example, the two parallel chains selected for a swap judgement no longer need to be constrained to adjacent chains; rather, they can be randomly selected from the entire set of chains.
Also, the number of swap judgements to be made at each random pause, that is, the number of pairs of chains selected in the parallel tempering process, can be set to arbitrary values other than 1 in the input file.
Section \ref{sec:eg} will demonstrate the different ways of controlling the parallel MCMC process using Nii-C code through a number of examples.

%

\section{Examples of applications}
\label{sec:eg} 

As Nii-C is designed to be a general-purpose MCMC code, it can be adapted to 
a variety of parallel sampling framework and a diverse range of applications.
To proceed, users can modify the controlling input parameter of Nii-C in order to regualte the parallel MCMC for a specific sampling strategy,
or write their own prior and likelihood functions in the template files ``\texttt{user\_prior.c}" and ``\texttt{user\_logll.c}" for a specific target distribution.
This section presents a number of applications and benchmark comparisions using the Nii-C code, a simple liner regression model, a hierarchical Bayesian model to infer the radius relationship of hot Jupiters, a planetary orbital parameter fitting model, a multimodal eggbox distribution, a curvedl likelihood model, a five-dimensional Rastrigin function, and a widely separated mixture of Gaussian distributions.

\subsection{Linear regression}

\begin{figure*}
\begin{center}
   \includegraphics[width=6.50in]{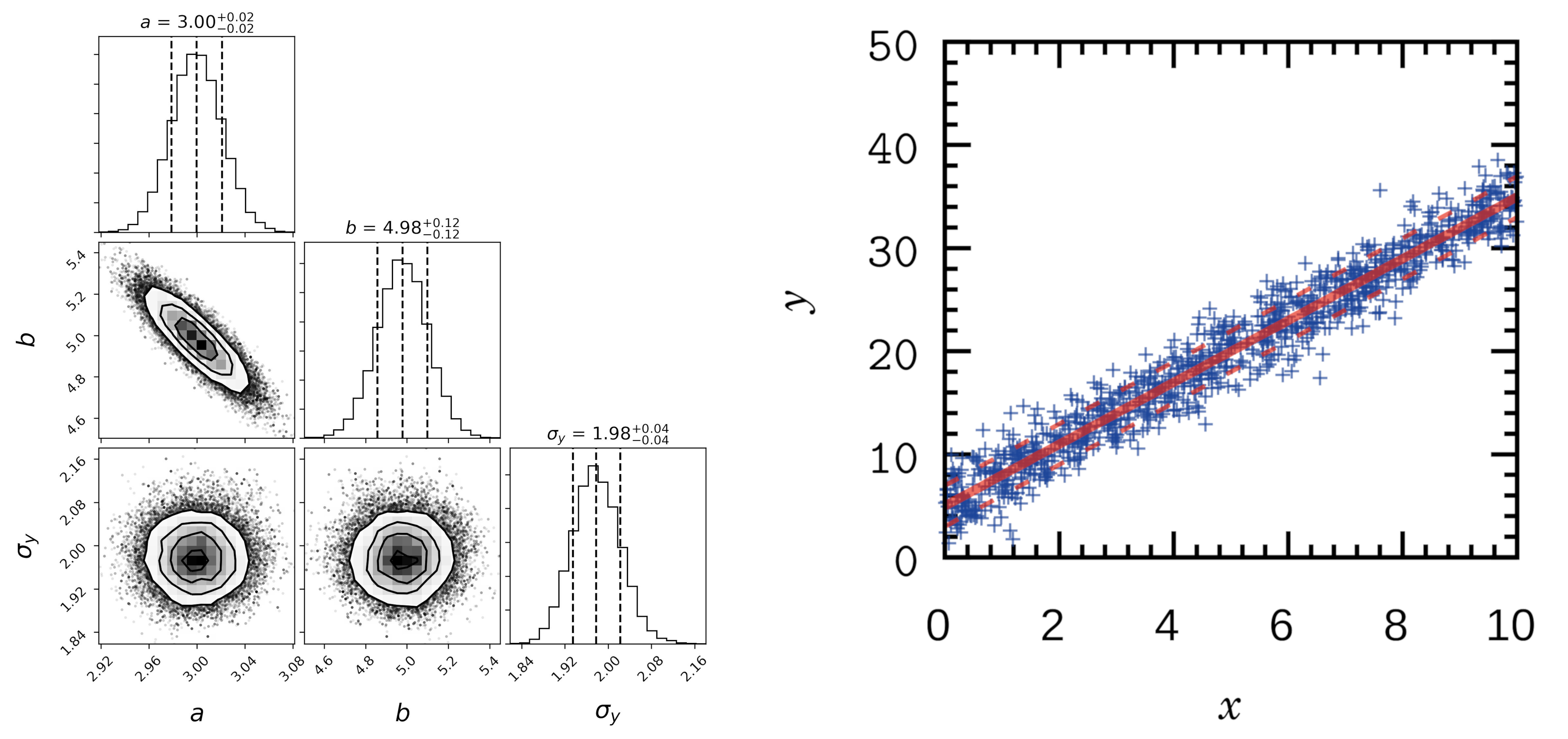}
\end{center}   
\caption{Nii-C fits a linear regression relationship to the simulated data points. 
The left panel shows the corner plots of the fit. The posterior means and one-sigma locations of the parameters $a$, $b$, and $\sigma_{y}$ are consistent with the original model that generated the data.
The right panel shows the simulated data as blue points and the linear relationship given by the posterior means of the fitted parameters as a red line. The red dashed lines indicate the  locations of a standard deviation according to the fitted $\sigma_{y}$.}
\label{fig:2}
\end{figure*}

A linear regression model that depicts the relationship between a dependent variable $y$ and an independent variable $x$ is a fundamental model used in data fitting.
Equation \ref{eq:reg} describes the linear regression model used in the Nii-C code to generate simulated data.
The model assumes the existence of additional Gaussian measurement errors described by a standard deviation of $\sigma_{y}$. 
A total of 1000 simulated $x$ and $y$ pairs were generated by combining three parameters, $a$, $b$, and $\sigma_{y}$, which were set to 3.0, 5.0, and 2.0, respectively.
The distributions of these data points in $x$ versus $y$ space are shown in the right panel of Figure \ref{fig:2}.

\begin{equation} \label{eq:reg}
    y \sim \mathcal{N}(ax+b, ~ \sigma_{y})
\end{equation}

For such a simple model, the parallel tempering algorithm is not necessary. 
The purpose of this test run is to demonstrate the functionality of the Nii-C code and to provide an easy-to-use template.
Two parallel tempering Markov chains were conducted for this linear regression model using Nii-C's APT-MCMC algorithm with beta values of 0.1 and 1.0, and each of the two Markov chains completed 90,000 iterations.
In this example, the parallel tempering MCMC was conducted without temperature ladders adjustment.
The first 10,000 iterations were set as the early tuning stage of proposal distributions and discarded as an initial burn-in.
The ranges of the prior distribution for the model parameters $a$, $b$, and $\sigma_{y}$ are set as  -20 to 20, -50 to 50, and 0.01 to 30 respectively. 
A ThinkPad X1 Carbon laptop was used to perform the corresponding APT-MCMC process.
The sampling process used 2 cores of the laptop's 8-core i7-8550U CPU.
The execution time was about 0.8 seconds. 
The corner plot in the left panel in Figure \ref{fig:2} shows the marginal posterior distributions of all model parameters.
The plot indicates that all three parameters effectively converge.
The sampled posterior means of $a$, $b$, and $\sigma_{y}$ are 3.00, 4.98, and 1.98, respectively, which is consistent with the parameters used to generate the data.

\subsection{Hierarchical radius relationship}

\begin{figure*}
\centering
   \includegraphics[width=5.5in]{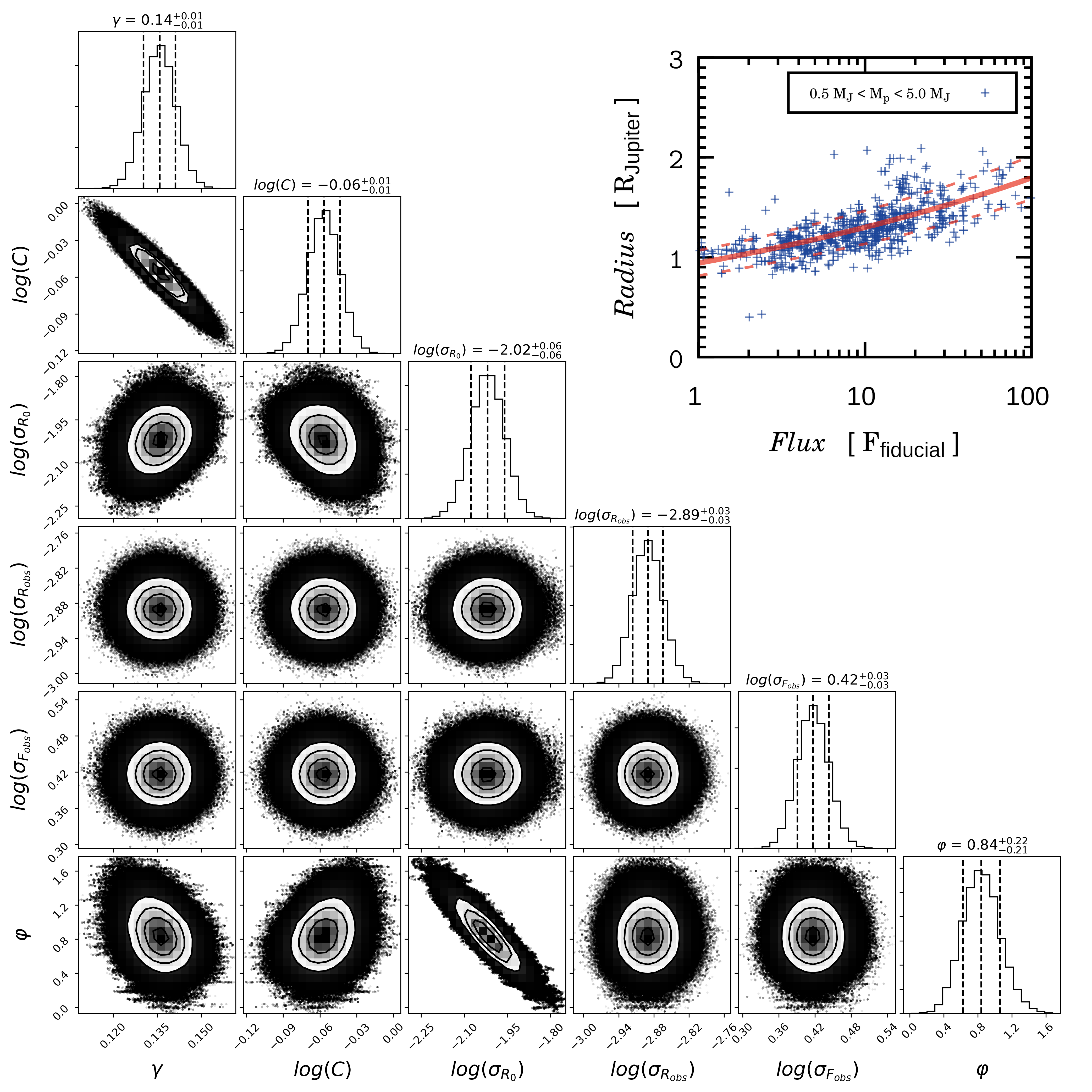}
\caption{
Corner plot of the marginal distributions of the fitted parameters in the hierarchical radius relationship model.
The distribution of hot Jupiters in the planetary radius versus incoming flux space is shown in the top right panel, along with the fitted power-law relationship (red line) and the locations corresponding to one standard deviation (red dashed lines).
}
\label{fig:3}
\end{figure*}

Hierarchical modeling is a powerful technique of Bayesian statistics.
In this technique, statistical models are formulated at multiple levels to account for dependencies between parameters, that is, the prior distribution of some model parameters is dependent on other parameters \citep{Lambert2018}. 

Many observed astrophysical quantities are subject to multiple layers of dependence. 
Therefore, accurate estimation of them requires a hierarchical Bayesian model to disentangle the interdependences \citep{Kelly2007,Loredo2019}.
One such example is the radius relationship of exoplanets obtained from transit observations \citep{Borucki2011,Batalha2013}.
The observed radii of exoplanets are influenced not only by physical quantities such as planetary mass and stellar flux, but also by the bias introduced by transit observations.
Hierarchical Bayesian models can be used to effectively fit this type of multi-level dependency data \citep{Wolfgang2016,Chen2017,Sestovic2018,Ning2018,Ma2021}.

The ``\texttt{hier\_R}" directory in Nii-C's source code provides a hierarchical Bayesian model explaining the relationship between planetary radius and the incoming stellar flux of hot Jupiters.
This hierarchical model consists of two levels.
At the first level, it assumes that both the dependent variable $R$ (planetary radius) and the independent variable $F$ (incoming stellar flux) have a Gaussian error distribution.
Accordingly, the statistical distributions of planetary radii and stellar flux can be expressed by the following two equations:

\begin{equation}
R_{\rm{obs}} \sim\mathcal{N}(R, ~~ \sigma_{R_{\rm{obs}}})
    \label{eq:Robs}
\end{equation}
\begin{equation}
F_{\rm{obs}} \sim\mathcal{N}(F, ~~ \sigma_{F_{\rm{obs}}}) 
    \label{eq:Fobs}
\end{equation}

where $R_{\rm {obs}}$ represents the observed transit radius of a hot Jupiter,  $F_{\rm{obs}}$ represents the calculated incoming stellar flux based on the observed quantities of a planetary system,  and $\sigma_{R_{\rm{obs}}}$ and $\sigma_{F_{\rm{obs}}}$ refer to the standard deviations of the normal distributions of the observation error. 

Note that the relationships given by Equaiton \ref{eq:Robs} and \ref{eq:Fobs} are only rough approximations, and in reality the Gaussian errors should be modelled independently for each planet and for each observation.
Such a simplification is based on the assumption that all observational errors for the entire hot Jupiter population can be modelled as a single Gaussian distribution \citep{Sestovic2018}.
In cases where the observational error is not large compared to the intrinsic dispersion of hot Jupiters, this approximation can effectively separate the two levels of Gaussian dispersion and help us describe the radius relationship more accurately.
A detailed analysis will be provided in a follow-up paper that aims to trace the sources of the inflated radii of hot Jupiters from a data perspective.

The incoming stellar flux $F$ can be calculated using:
\begin{equation}
F = \left( \frac{R_{\rm *}}{a} \right)^2 \sigma T_*^4 \label{eq:incident-flux}
\end{equation}
where $a$ is the orbital semi-major axis, $\sigma$ is the Stefan–Boltzmann constant, $T_{\rm *}$ and $R_{\rm *}$ denotes the temperature and radius of the host star.
Additionally, $F$ is standardized to 1 at a distance of 0.1 AU from a solar-like star having an effective temperature of 5777 K.

The second level in the hierarchical model is the statistical relationship between planetary radii and  incoming stellar flux.
The model assumes that the planetary radius at a given incoming stellar flux follows a Gaussian distribution.
Furthermore, it also assumes that the standard deviation of the Gaussian radius dispersion can vary as a function of radius, following a slope denoted as $\varphi$.
Such a relationship between planetary radius and incoming stellar flux can be expressed as follows:
\begin{equation}
R \sim \mathcal{N}(C F^{\gamma}, ~~ \sigma_{R_{\rm{0}}}\cdot(C F^{\gamma})^{\varphi})\label{eq:12} 
\end{equation}
where $C$ and $\gamma$ are a constant and a power-law index in the radius relationship, respectively, $\sigma_{R_{\rm{0}}}$ is a reference standard deviation, and $\varphi$ is a power-law index that denotes the change in standard deviation across different planetary radii.

For prior distributions, $\gamma$ is modeled as a normal distribution, the other five model parameters,  $\varphi$, $\ln({C})$, $\ln(\sigma_{R_{\rm{obs}}})$, $\ln(\sigma_{F_{\rm{obs}}})$, and $\ln(\sigma_{R_{\rm{0}}})$, are modeled as uniform distributions.

To sample the hierarchical model, the Metropolis MCMC is run in the parameter space together with $R$, based on the posterior probability values determined by the hierarchical radius relations that describe all the observed data of hot Jupiters.
The observational data for all hot Jupiters are downloaded from the NASA Exoplanet Archive \footnote{\url{https://exoplanetarchive.ipac.caltech.edu}}. 
The top right panel in Figure \ref{fig:3} shows the distribution of all these hot Jupiters in the planetary radius versus incoming flux space. 
Using this data, the hierarchical radius relationship model is sampled with the Nii-C code using four parallel tempering chains with $\beta$ values of 0.001, 0.01, 0.1, and 1.0, 
with the temperature ladders adjustment switched off.
In total, 10,000,000 iterations were sampled, of which the first 1,000,000 iterations were set as the tuning stage and discarded as an initial burn-in.
The entire sampling process of the 10,000,000 iterations was divided into 2000 stacks of 5000 iterations each.
After the first few stacks in the early tuning stage, the proposal distributions of all model parameters are adjusted using Algorithm \ref{alg:one} for the chains that have obtained poor acceptance rates.
The execution time of this run is approximately 1202.4 seconds on the identical ThinkPad X1 Carbon laptop with an i7-8550U processor.

Figure \ref{fig:3} displays the corner plot of the marginal posterior distributions of all the parameters in the hierarchical model. 
The posterior means of $\gamma$, $\varphi$, $\ln({C})$, $\ln(\sigma_{R_{\rm{obs}}})$, $\ln(\sigma_{F_{\rm{obs}}})$, and $\ln(\sigma_{R_{\rm{0}}})$ are 0.14, 0.84, -0.06, -2.89, 0.42, and -2.02, respectively.
The power-law relationship between planetary radius and incoming stellar flux given by the fitted parameter values is plotted in the top right panel in Figure \ref{fig:3}.
The fitted curve clearly shows that both the radii of hot Jupiters and the standard deviation of their radius distributions increase with the incoming flux, in agreement with the observation.

\subsection{Orbital parameter fitting}

\begin{figure*}
\centering
   \includegraphics[width=6.5in]{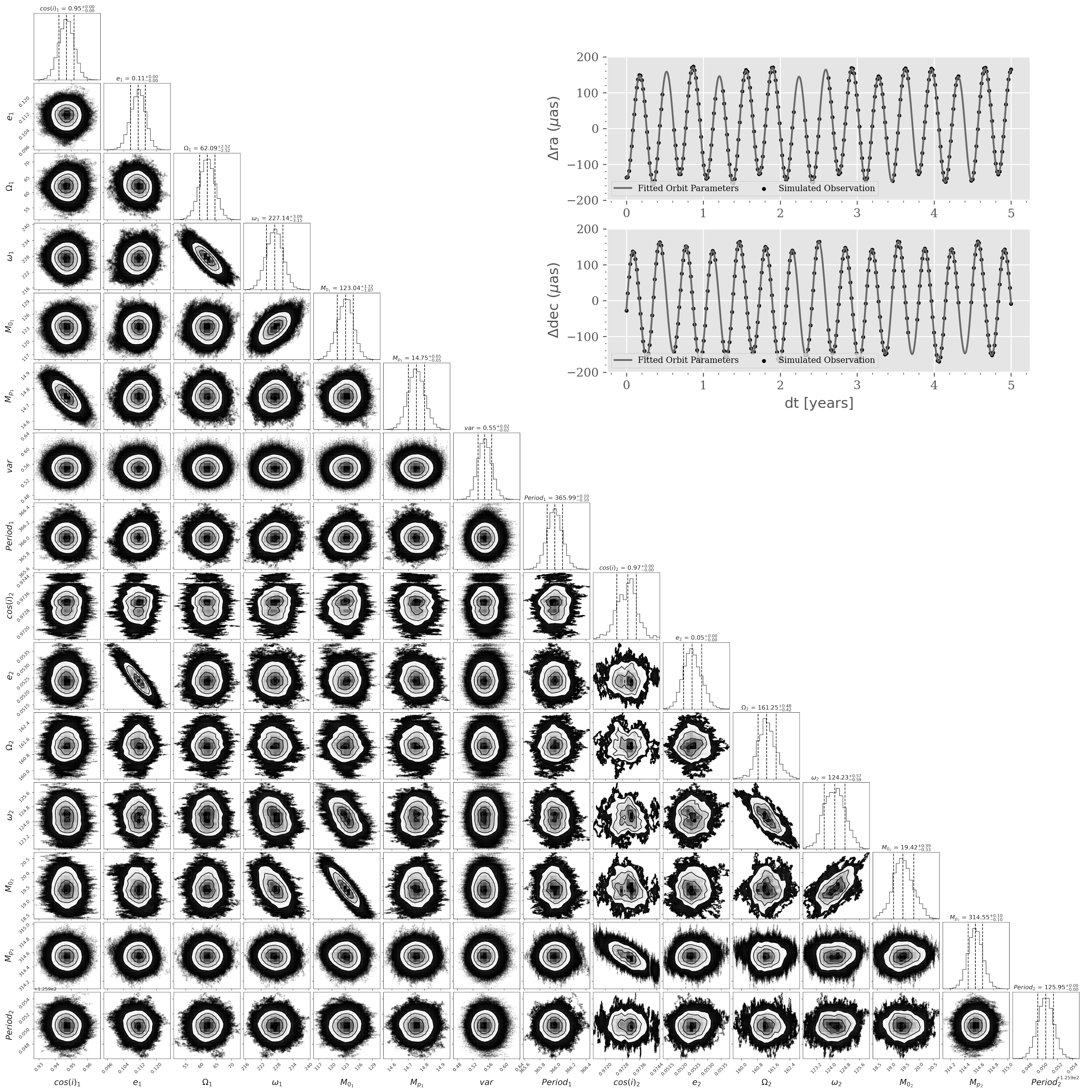}
\caption{Corner plot of the marginal distributions of all 15 parameters in the orbit fitting of a two-planet system.
The top right panel shows the comparision between the simulated observation and the theoretical signals generated using the fitted parameters.
    }
\label{fig:4}
\end{figure*}

Nii-C's convergence capability and fast execution speed enable efficient sampling of complex distributions.
One such application is a blind search for the orbital parameters of a two-planet system based on the astrometric measurements of a star.
Although Bayesian inference is an effective approach for determining the orbital parameters of a planetary system around a star \citep{Gregory2005b,Balan2009,Gregory2011,Schulze2012,Perryman2014,Wu2016,Feng2019,Blunt2020,Brandt2021,Wu2023}, the difficulty increases with the number of planets because the number of model parameters is proportional to the number of planets.
Using the Nii-C code, the fitting of the orbital parameters of a mutiple planet system can be achieved in a short execution time.
The ``\texttt{fit\_orbit}" directory in the source code contains an example of fitting the 15 orbital parameters of a two-planet system using simulated astrometric observations.
In this two-planet system, there is a 314.71 $M_{\oplus}$ Jovian planet with an orbital period of 125.95 days and a 14.71 $M_{\oplus}$ planet with an orbital period of 365.95 days.
Other orbital parameters of this two-planet system are listed in Table \ref{tab1}.
The astrometric signals of this two-planet system were generated using a Python code that simulate differential astrometry observations \citep{Jin2022}.
The prior and likelihood functions are also set based on differential astrometry \citep{Jin2022}.

\begin{table}
\caption{The true values of all 15 model parameters and the fitted posterior means of the two-planet model.}
 \centering
\begin{tabular}{ccc}
\hline
\hline
Parameter & True Value  & Posterior Mean \\ 
\hline
$P_1 ({\mathrm {days}})$&  365.95  & 365.99  \\
$P_2 ({\mathrm {days}})$&  125.95 &  125.95  \\
$M_{{\mbox{p}}_1} (M_{\oplus})$ & 14.71 & 14.75  \\
$M_{{\mbox{p}}_2} (M_{\oplus})$ & 314.71 & 314.55  \\
$e_1$ &  0.10  &  0.11  \\
$e_2$ &  0.05 &  0.05 \\
$cos~i_1$ & 0.95 & 0.95  \\
$cos~i_2$ & 0.97 & 0.97  \\
$\Omega_1$ &  61.19 & 62.09  \\
$\Omega_2$ &  161.19 & 161.25  \\
$\omega_1$ & 224.83 &  227.14 \\
$\omega_2$ & 124.83 &  124.23  \\
$M_{{\mathrm 0}_1}$ & 120.01 & 123.04  \\
$M_{{\mathrm 0}_2}$ & 20.01 & 19.42  \\
$\epsilon_{x} (\mu\mbox{as})$ & 0.50 & 0.55 \\
\hline
\end{tabular}
\label{tab1}
\tablecomments{Subscripts 1 and 2 are used to distinguish between two different planets.}
\end{table}

The posterior probability distribution of the astrometric signals of the two-planet model is sampled with the Nii-C code using eight parallel tempering chains with $\beta$ values of 0.01, 0.03, 0.1, 0.2, 1.0, 0.4, 0.1, and 0.05.
These temperature ladders was set based on our previous experience of planetary orbit fitting.
Consequently, the option to adjust the temperature ladders of parallel tempering MCMC was disabled.
A total of 10,000,000 iterations were sampled, with the first 3,000,000 iterations were set as the tuning stage and discarded as initial burn-in.
The sampling process takes 3471.7 seconds using the same ThinkPad X1 Carbon laptop with an i7-8550U processor.
Table \ref{tab1} summarises the posterior means of the 15 model parameters.
It shows that all the true values of model parameters were accurately retrieved.
Figure \ref{fig:4} gives the corner plot of the marginal posterior distributions of all 15 parameters in the two-planet system.
The top right panel in Figure \ref{fig:4} compares the simulated observation and the theoretical signals generated using the fitted parameters.
It clearly shows that the astrometric signals from both the shorter-period Jovian planet and the long-period low-mass planet are well fitted.
To ensure convergence of MCMC, additional independent APT-MCMC runs were performed using different sets of initial random seeds. 
The independent runs show that chains with $\beta$ values equal to 1 converge to the same parameter interval, indicating that convergence has been achieved.

\begin{figure*}
\centering
   \includegraphics[width=6.5in]{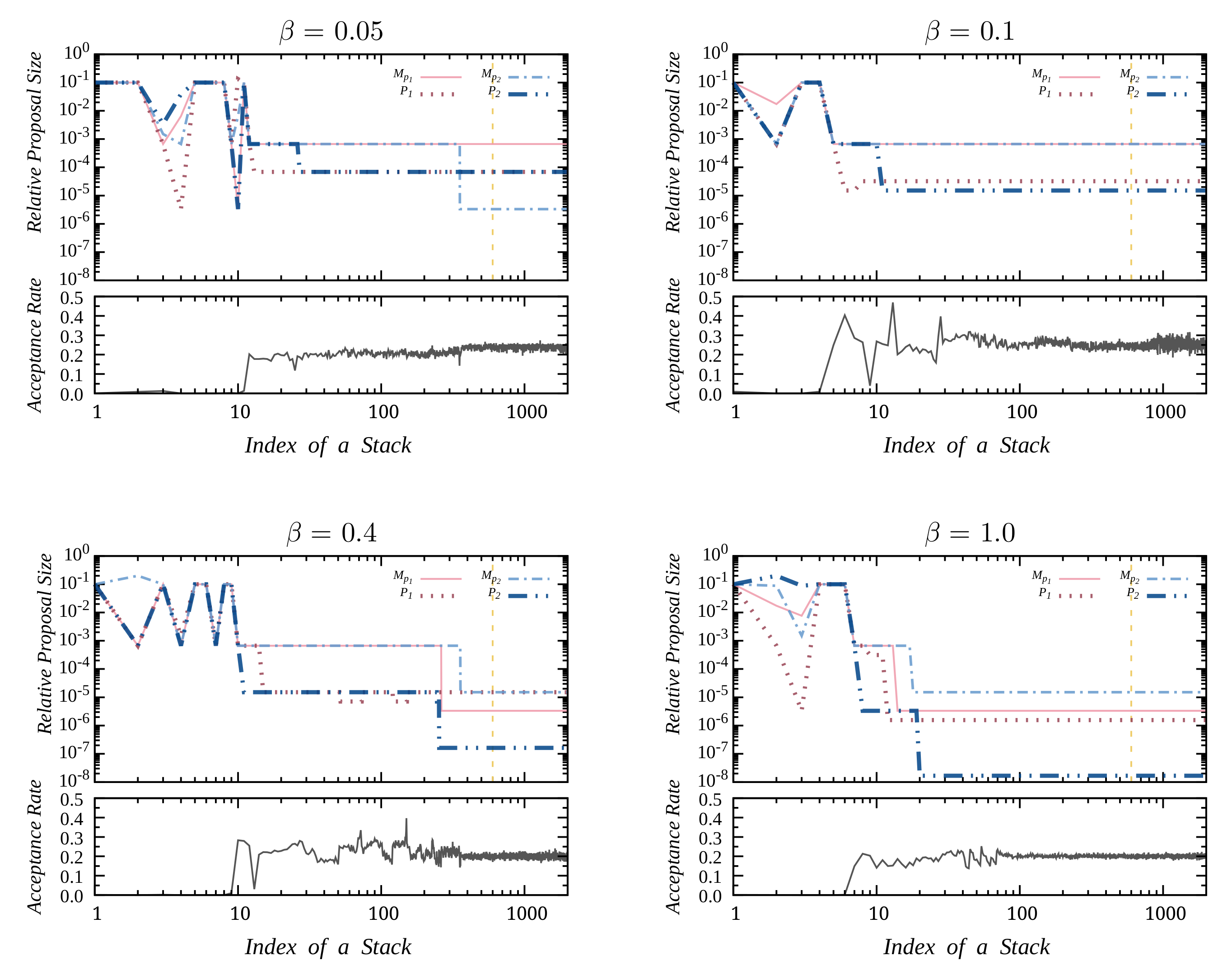}
    \caption{The relative sizes of the proposal distributions of four model parameters (planetary masses and orbital periods) and the resulting acceptance rates for each of the 2000 batches in four neighboring parallel tempering chains with $\beta$ values equal to 0.05, 0.1. 0.4 and 1.0.
    Note that for the chain with a $\beta$ of 1.0, the mass and period of the first planet have similar relative proposal sizes, while the relative proposal sizes of the mass and period of the second planet differ by a factor of 1000.
    Nii-C can efficiently determine the optimal proposal sizes for all model parameters.
    The end of the tuning stage is indicated by the vertical dashed lines.}
\label{fig:5}
\end{figure*}

The entire sampling process of the 10,000,000 iterations was divided into 2000 stacks of 5000 iterations each.
To track the behavior in the tuning stage, Figure \ref{fig:5} shows the relative sizes of Gaussian proposal distributions for planetary masses and orbital periods in four parallel chains with $\beta$ values of 0.05, 0.1, 0.4 and 1.0, along with the corresponding sampling acceptance rates in these four neighbouring chains.
The remaining eleven model parameters and the other four  parallel chains with different $\beta$ values are not shown in the graphs for brevity.
The standard deviations of the Gaussian proposal distributions for all parameters are initially set to 0.1 times the ranges of their respective prior distributions.
Such large proposals allow all parallel tempering chains to efficiently explore the global posterior distribution.
Subsequently, each chain resets its sampling proposals based on the acceptance rate of the previous stack, if the rate is unsatisfactory. 
Thus, a change in the proposals means that the sampler of a chain starts exploring a new region in the posterior distribution, where the previously used proposals are not suitable for the morphology of this region.

The purpose of tuning the Gaussian proposals for all the parallel tempering chains is to maintain an appropriate acceptance rate.
Consequently, as shown by the acceptance rates of the four neighbouring chains in Figure \ref{fig:5}, changes in the size of the proposals are accompanied by a jump in the acceptance rate, as the proposals for a chain are tuned whenever the acceptance rate deteriorates.
Therefore, by examining the resulting acceptance rates of each chain, it is possible to infer changes in the proposals of the other 11 model parameters, even though Figure \ref{fig:5} only displays their evolution. 
For the four chains with $\beta$ values of 0.05, 0.1, 0.4 and 1.0, the proposals of all 15 model parameters remained stable after the first 353, 29, 358, and 69 stacks, suggesting that these four chains reach convergence and do not deviate from it thereafter.
Note that the option to adjust the proposals is no longer available after the first 600 stacks to maintain the Markovian property in parallel tempering chains.
This setting does not affect the final sampling results since the late sampling period is typically very stable.

Intuitively, the chain with a lower $\beta$ value is expected to be more stable because the distribution is flatter compared to the case with a larger $\beta$.
This is the general trend observed from the changes in the proposals of the four neighbouring chains shown in Figure 5, although it is not strictly followed.
The stability of a chain is affected not only by its own $\beta$ value, but also by the $\beta$ values of the adjacent parallel tempering chains.
If a chain adjusts its sampling proposals, it is likely to trigger its neighbouring chains to adjust their proposals.
Therefore, it is important to have a suitable set of ladder-like $\beta$ values to ensure fast convergence of a distribution, particularly for complex models.
In the future, Nii-C's efficiency can be further improved by adding the option to use a dynamic combination of $\beta$ values \citep{Vousden2016}.

In an APT-MCMC process, it is important to maintain appropriate sampling acceptance rates for the chains with a $\beta$ value of 1.
These chains correspond to the original form of the posterior distribution of a model.
The bottom right panel of Figure \ref{fig:5} plots the acceptance rates of the chain with a $\beta$ = 1 in the Beyasien orbit fitting of the two-planet system.
From the eighth stack onwards, this chain maintains an acceptance rate of around $\sim$ 0.2 as the sampler moves towards convergence.
This explains Nii-C's extraordinary ability to converge.

Note that we utilize the blind search method to fit orbital parameters  only for the evaluation of the performance of the Nii-C code.
Subsequently, we will optimise the orbit fitting schedule using a model that can reduce the orbital parameters of a planet to 3 \citep{Catanzarite2010}, further reducing the execution time of the orbital retrieval.

\subsection{Eggbox distribution}

\begin{figure*}
\centering
   \includegraphics[width=6.5in]{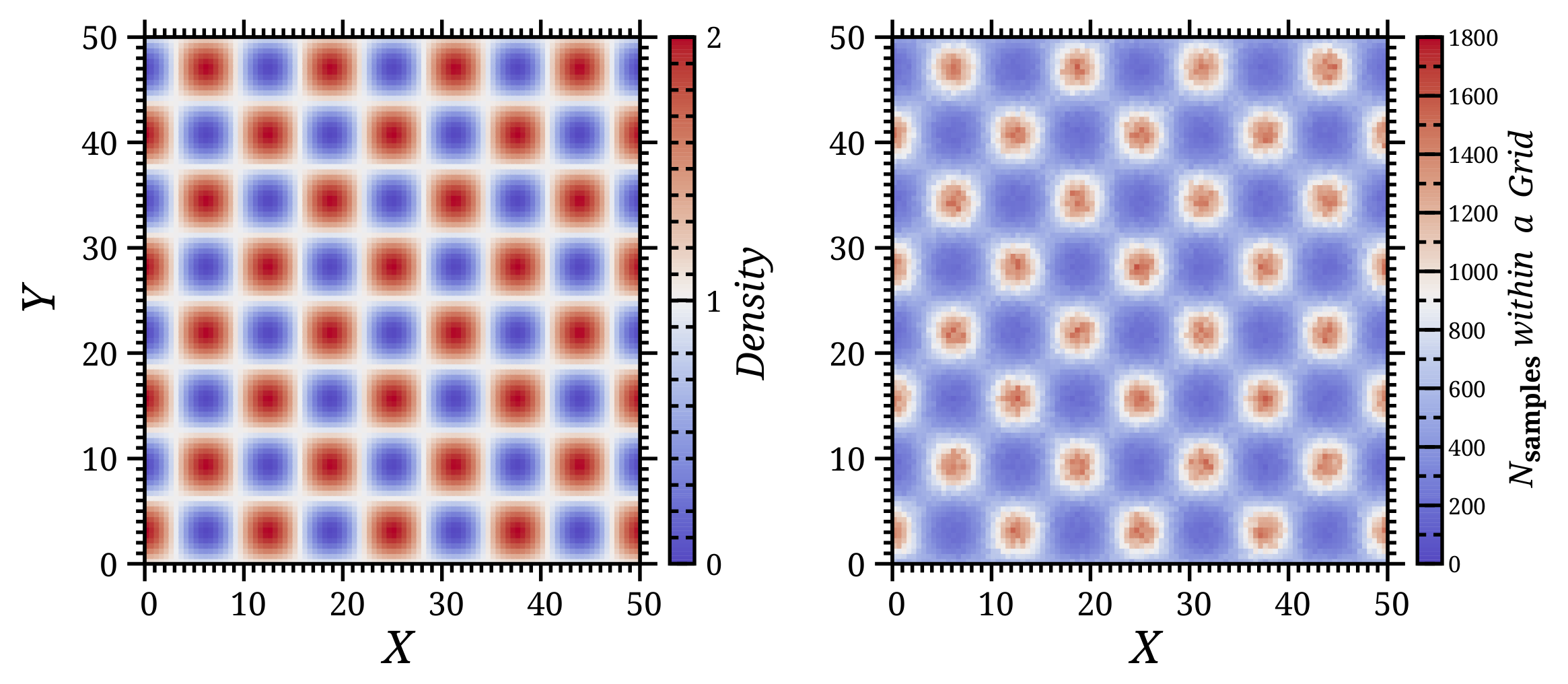}
\caption{
The left panel displays the theoretical distribution given by Equation \ref{eq:eggbox}.
The right panel shows the  number of effective sampling points obtained by 15 walkers using Nii-C, configured as a basic parallel MCMC code, in each of the 250 grids in the $X$-$Y$ plane.
}
\label{fig:6}
\end{figure*}

An Eggbox function is a surface obtained by combining a sine and a cosine function perpendicularly.
It is a typical multimodal distribution that is characterised by a large number of local extrema and saddle points.
The ``\texttt{eggbox}" directory in the source code presents an implementation of an Eggbox distribution using Nii-C based on Equation \ref{eq:eggbox}.

\begin{equation}
Eggbox(x,y) = cos(x/2)*sin(y/2) + 2 \label{eq:eggbox}
\end{equation}

To demonstrate Nii-C's ease of adaptation to different sampling strategies, we turned off the auto-adjustment of temperature ladders and proposal distributions in Nii-C, and set all the $\beta$ values of parallel chains to 1, making Nii-C a basic parallel MCMC code.

A total of 15 workers were used in the experiment, with each worker sampling 405,000 iterations.
The intial 5000 iterations were discarded as initial burn-in, resulting in a  total of 6,000,000 effective iterations for both runs.
Figure \ref{fig:6} presents the model eggbox distribution and the sampling results obtained by Nii-C.
The sampling process takes 0.154 seconds using a ThinkPad X1 Carbon laptop with an i7-1360P CPU.
The experiment demonstrates that Nii-C can be configured as a basic parallel MCMC sampling code and that it performs well in practice.

\subsection{Curved likelihood}

\begin{figure*}
\centering
   \includegraphics[width=6.5in]{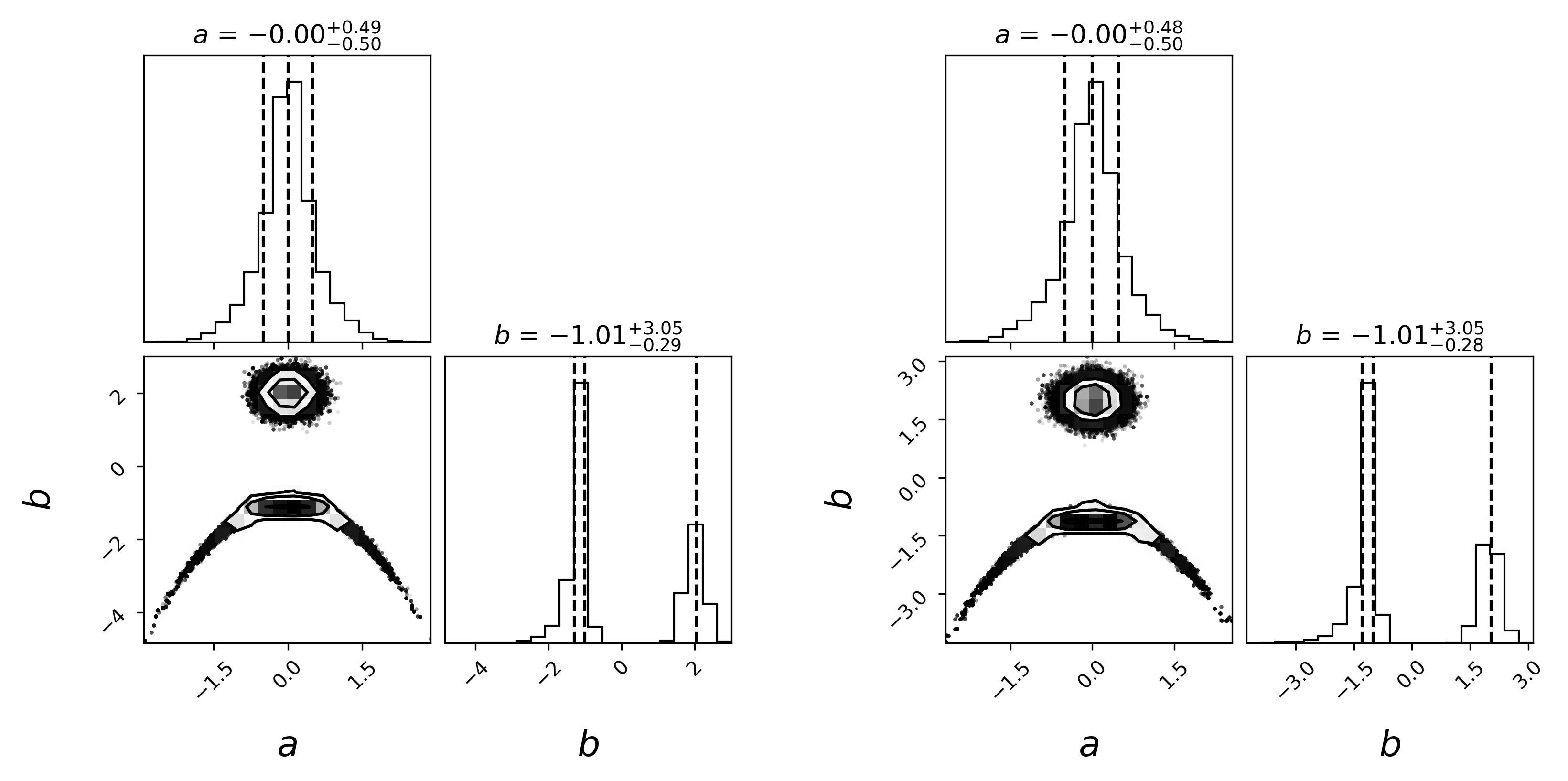}
\caption{
    The left panel shows the corner plots of the sampling results of the curved likelihood obtained by Nii-C, while the right panel shows the results obtained by PTMCMCSampler.
    }
\label{fig:7}
\end{figure*}

In this benchmark, Nii-C is compared with the Python language parallel tempering MCMC sampler PTMCMCSampler \citep{Ellis2019}.
As PTMCMCSampler is a standard parallel tempering MCMC code, we have turned off Nii-C 's automatic tuning of temperature ladders and proposal distributions of to make it sample according to standard parallel tempering MCMC.
It should be noted that the comparison presented here is solely intended to contrast the sampling outcomes and computational efficiency of the two codes associated with parallel tempering MCMC.
For better sampling effect in real-world applications, it is recommended that the Nii-C code be configured to allow the automatic tuning of the proposal distributions during burn-in, which is disabled in this instance.

\begin{equation}
exp(-x^2 - (9+4x^2 + 9y)^2) + 0.5 \times exp(-8x^2 - 8(y-2)^2 ) \label{eq:curvedlh}
\end{equation}

The test distribution is a curved likelihood that provided in examples of the PTMCMCSampler source code, which is depicted by Equation \ref{eq:curvedlh}.
In both runs, four parallel tempering MCMC chains with $\beta$ values of 0.03, 0.1, 0.3 and 1.0 are set  using either PTMCMCSampler or Nii-C, run on a ThinkPad X1 Carbon laptop with an i7-1360P CPU.
To compare the two codes, we performed five identical runs and calculated the average execution time.
The average runtime of Nii-C is 0.392 seconds, while that of the PTMCMCSampler is 45.889 seconds.
As can be seen in Figure \ref{fig:7}, the two codes give similar results, except that PTMCMCSampler is on average 117.06 times slower than Nii-C.
The two orders of magnitude speed difference between PTMCMCSampler and Nii-C is largely attributable to the disparity in computational efficiency between Python and C, since the mpi4py library requires a significant amount of computation time compared with the MPI library in C \citep{Asaduzzaman2024}.
Moreover, in real applications the computational cost of likelihood calculation will be dominant, and a Python user can make use of C extensions or other vectorization techniques to make the likelihood calculation in PTMCMCSampler to be comparable to C. 
The different characteristics of these two codes make each of them more suitable for different applications, and our comparison only serves to illustrate the excellent computational efficiency of Nii-C.
The high computational efficiency of Nii-C suggests that it can serve as an ideal platform for parallel sampling.

\subsection{Five-dimensional Rastrigin function}

\begin{figure*}
\centering
   \includegraphics[width=6.5in]{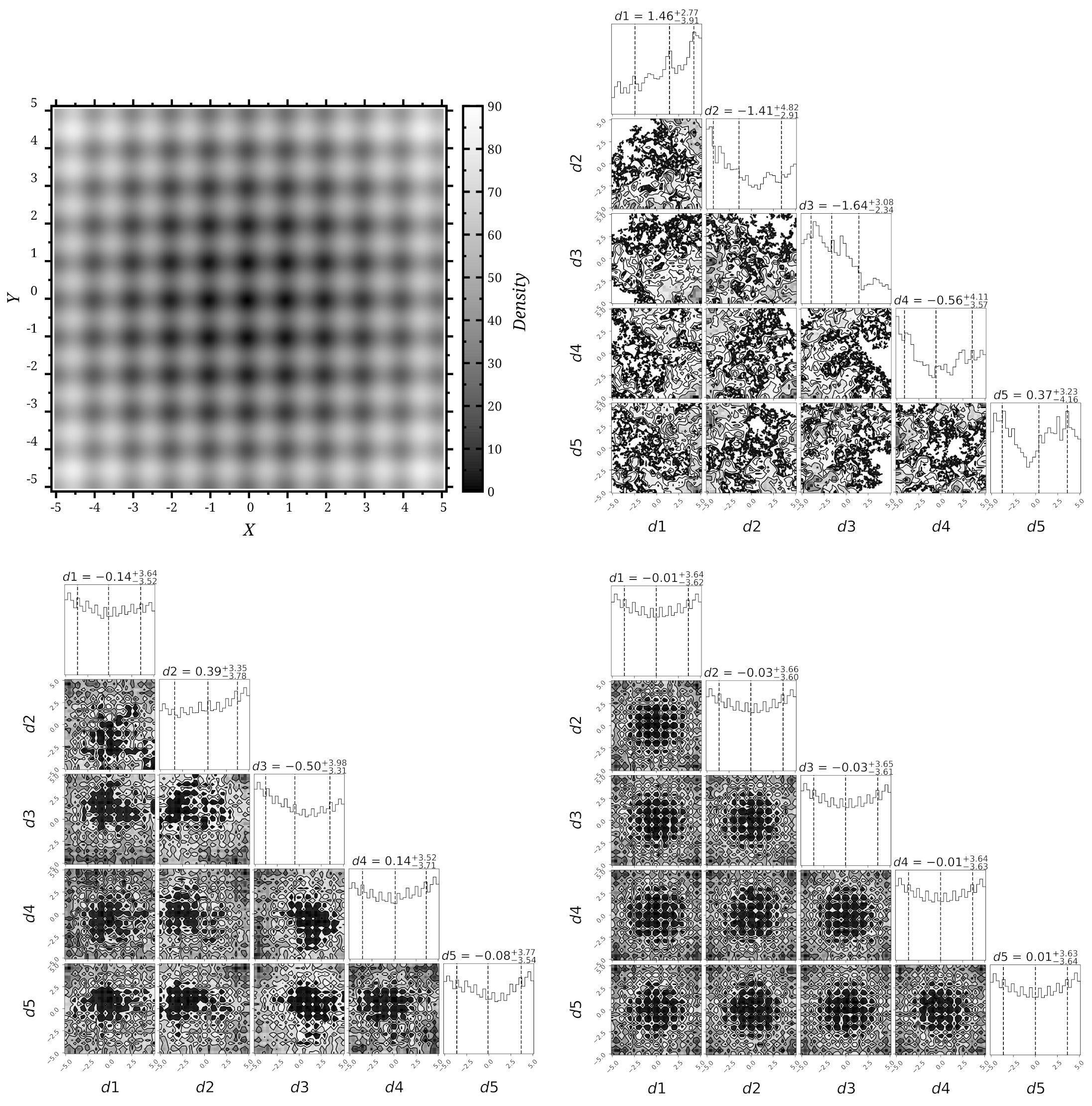}
\caption{
The sampling outcomes of a five-dimensional Rastrigin function using Nii-C with different configurations.
The top left panel gives the theoretical distribution of a two-dimensitional Rastrigin function.
The top right panel displays the corner plots resulting from the implementation of standard parallel tempering MCMC configuration, whereby the early tuning of the pallallel tempering ladders and proposal distributions have been disabled.
The bottom left panel displays the results obtained in the case where only the early tuning of proposal distributions proposal is enabled,  while the tuning of parallel tempering ladders is disabled.
The bottom right panel shows the case in which both the parallel tempering ladders and proposal distributions proposal are activated in the initial tuning stage.
    }
\label{fig:8}
\end{figure*}

This sub-section demenstrates the effects of early tuning of parallel tempering ladders and proposal distributions  using a five-dimensional Rastrigin function.
The generalized Rastrigin function is a non-convex function employed as a performance test problem for optimization algorithms \citep{Muhlenbein1991}.
It represents a typical example of a non-linear multimodal function with  a vast search space and a large number of local minima.
The Rastrigin function on an $n$-dimensional domain is defined by:
\begin{equation}
    f(\mathbf{x}) = An + \sum_{i=1}^{n} \left[x_i^2 - A cos(2\pi x_i) \right] \label{eq:rastrigin}
\end{equation}
where $A = 10$ and $x_i \in [-5.12, 5.12]$.

In the case of the five-dimensional Rastrigin function, standard parallel tempering MCMC sampling method works well when the proposal distributions of all parameters are set to one tenth of their parameter distributions, paired with logarithmically uniform ladders.
In order to demonstrate the effectiveness of Nii-C's automatic tuning system, we deliberately set up a bad initial configuration of  parallel tempering MCMC for three test runs.
In all three runs, the initial scale of the proposal distribution for each parameter is set to one percent of its respective distribution range.
The parallel temperature ladders were set to 1000, 500, 333.3 and 1, corresponding to $\beta$ values of 0.001, 0.002, 0.003 and 1.
All three test runs sample 2,000,000 iterations of the five-dimensional Rastrigin function using Nii-C, with the first 1,000,000 iterations discarded as initial burn-in.
They are characterised by their statuses as a standard parallel tempering MCMC, a parallel tempering MCMC with automatic tuning of proposal distributions, and a parallel tempering MCMC with automatic tuning of both the proposal distributions and the temperature ladders, respectively.

Each the three runs was completed in a few seconds using a ThinkPad X1 Carbon laptop with an i7-1360P CPU.
Figure \ref{fig:8} displays the sampling results of the three comparative runs, with the two-dimensional Rastrigin function shown as a reference.
As can be seen in the top right panel, selecting an appropriate combination of proposal distributions is of paramount importance for achieving a reasonable sampling outcome.
In the case of a bad combination of proposal distributions, standard parallel tempering MCMC is completely unable to obtain useful results, indicating that the parallel samplers are trapped in local extremes.

The bottom left panel of Figure \ref{fig:8} shows the sampling outcome of the parallel temperaing MCMC with automatic tuning of proposal distributions.
This sampling results can be used to probe the overall distribution characteristics of the five-dimensional Rastrigin function, except that the distribuion obtained deviates significantly from the exact result, suggesting that the parallel temperature ladders are not optimal.
This is in contrast to the completely useless result obtained in the top right panel of a standard parallel tempering MCMC, demonstrating the importance of automatically tuning the proposal distributions at an early stage.
Especially for the majority of practical applications, where most probability densities are distributed over relatively narrow parameter intervals, and where the relative distribution width of each parameter varies considerably (For example, the relative proposal sizes of different parameters shown in Figure \ref{fig:5}), dynamically adjustment of the proposal distributions  at an early stage determines whether or not a valid sampling results can be obtained.

The bottom right panel of Figure \ref{fig:8} displays the sampling result of the case that allows automatic tuning of both the proposal distributions and the temperature ladders.
As can be seen from the two-dimensional distributions of each pair of parameters and the marginal distribution of each parameter, the result of this run are in accordance with the true distribution of the five-dimensional Rastrigin function.
In the burn-in stage of this run, the first 300,000 iterations are set to automatically tune the temperature ladders, and the later 700,000 iterations are set to tune the proposal distributions. 
In comparison, the bottom left run did not do  automatic tuning for the temperature ladders during the first 300,000 iterations.
When the ladders tuning is enabled, the $\beta$ values are tuned from 0.001, 0.002, 0.003 and 1 to 0.001, 0.014, 0.218 and 1, resulting in a significant improvement in the sampling outcome.

\subsection{Multimodal Gaussian}

\begin{figure*}
\centering
   \includegraphics[width=6.5in]{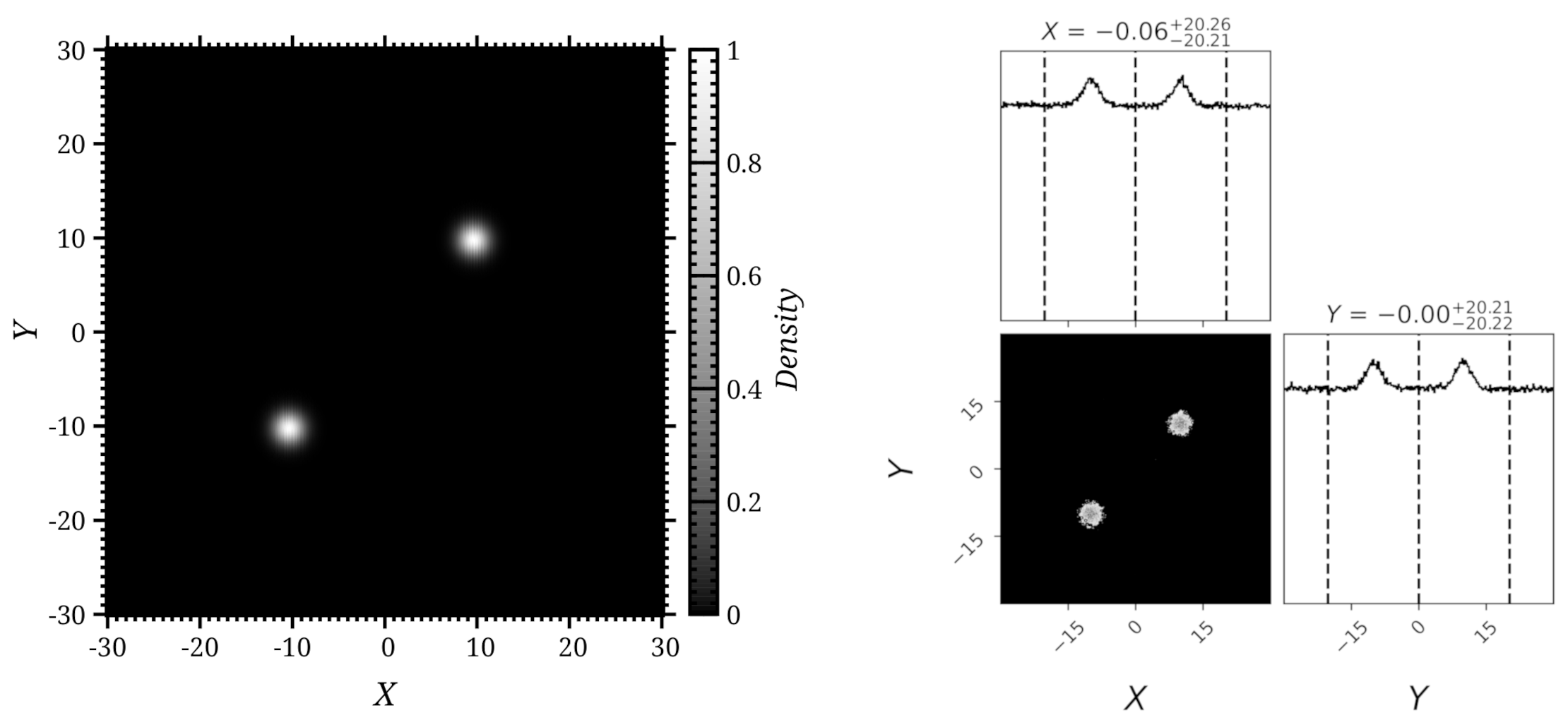}
\caption{
A multimodal two-dimensional Gaussian  distribution, left panel, and the results of an APT-MCMC sampling using Nii-C, right panel.
The standard deviations of the two peaks of this multimodal Gaussian distribution are both of equal to 1 on each axis.
The centres of the two peaks are located at (10,10) and (-10,-10), respectively.
    }
\label{fig:9}
\end{figure*}

In this example, we test a situation where the standard MCMC algorithm does not perform well, that is, multiple peaks that are separated by a large distance.
The left panel of Figure \ref{fig:9} shows the model distribution.
It is a two-mode Gaussian distributions with a standard deviation of 1 on each axis for each modes, and these two modes are separated by a distance of $\sim$ 14.14 times the standard deviation.
Using Nii-C, this distribution is sampled for 5,000,000 iterations using an APT-MCMC procedure.
The first 500,000 iterations were set as the early burn-in phase, which was further divided into an earlier 300,000 iterations for auto-tuning the temperature ladders and a later 200,000 iterations for auto-tuning the proposal distributions.
The sampling process takes 2.88 seconds using a ThinkPad X1 Carbon laptop with an i7-1360P CPU.

The right panel of Figure \ref{fig:9} shows the outcome of the APT-MCMC sampling.
It can be seen that the morphology of this multimodal Gaussian distribution is well reproduced.
The initial temperature ladders are set as $\beta$ = 0.0001, 0.001, 0.01, 0.1 and 1, using 5 parallel tempering chains.
During the automatic tuning of the temperature ladders, the $\beta$ values of the parallel tempering chains vary slightly because the swap acceptance ratios were relatively close for adjacent pairs.
Following the temperature tuning process, the resulting $\beta$ values were found to be 0.0001, 0.0010, 0.0104, 0.0973, and 1.0, respectively.
The results of the automatic tuning of the proposal distributions vary considerably depending on the temperature of each parallel tempering chain.
For the parallel tuning chain that samples the target distribution, the chain with $\beta$ = 1, the proposed distribution of each parameter is stable at 0.1 of the parameter range after tuning, despite variations in the tuning process.

%

\section{Discussion}

\label{sec:diss}

The advantage of the Nii-C code is its fast execution speed and its exceptional convergence capability.
Nii-C employs two mechanisms to achieve this.
First, it implements the parallel tempering algorithm to help each chain's sampler escape local extremes.
Second, it develops a control system with the ability to automatically adjust the parallel temperature ladders and the proposal distributions of all parameters in an early tuning stage.
Because Nii-C is written entirely in C, it has superior execution speed.
To facilitate the use of Nii-C in various models, user-specific functions such as prior and likelihood have been designed to be as self-contained as possible and separated into two distinct source files.

Furthermore, as Nii-C is an open source software, users can integrate various sampling algorithms, in addition to the basic Metropolis–Hastings algorithm provided in the code, to accommodate different sampling strategies.
For example, the Langevin algorithm \citep{Roberts1998} will be implemented in the future  version of the Nii-C code.
The parallel tempering control system in the Nii-C code can also be modified if necessary. This paper can be actually viewed as a concrete example application of a general philosophy, of doing any kinds of tunings automatically and, if possible, also in parallel, {\it ahead of time}, in order to win the race of total computation time at the end.  This is currently applied to tuning the scale of the random walk proposal used in parallel tempering, in order to achieve a good acceptance rate near 0.234. However, many other things could be done too, in terms of both what to tune and what is the goal of tuning. Other goals of tuning may include how fast the chain moves, or how high the posterior probabilities are. Other things to tune may include starting points of the Markov chain. 
For instance, a gradually increasing ladder-like numerical sequence of $\beta$ is crucial for achieving efficient convergence when sampling complex models.
The general idea of automatic pre-tuning may also be applied to pre-tune the higher-level tuning parameters of the step-size sequence  (the constant and the exponent) used in the updates of some adaptive MCMC methods \citep{Miasojedow2013}.
Also, in some situations, pre-tuning may aim to achieve an ideal acceptance rate that differs from 0.234, depending on the specific ideal acceptance rates of various MCMC algorithms \footnote{For a list of MCMC algorithms and their ideal acceptance rates, see, e.g., \url{https://m-clark.github.io/docs/ld_mcmc/index_onepage.html}}.

The Nii-C code has potential applications in various research fields due to its efficiency in achieving convergence. 
This is particularly relevant in modern scientific researches  where complex models and large volumes of data pose significant challenges in Bayesian reference. 
In the future, Bayesian reference applications for diverse research topics will be continuously developed and added into the source code of Nii-C, offering researchers with efficient solutions for various research topics.

\appendix

\section{Source files in the Nii-C code}

The structure of the Nii-C code has been designed to facilitate its utilization in a wide range of applications.
Nii-C separates the source files dependent on a user's model, which correspond to the prior and likelihood functions of the model, from the bulk program subroutines that are linked to the adaptive parallel tempering MCMC.
This following list describes the primary subroutine files of Nii-C.

\begin{itemize}
   \item  \texttt{main.c} $-$ This file contains the main function of the program.  It is responsible for configuring and executing the entire APT-MCMC process.

   \item \texttt{mpi\_init.c} $-$ It initializes the random values of all model parameters in each  parallel tempering Markov chain.

   \item  \texttt{mpi\_flow.c} $-$ This file is responsible for scheduling tasks throughout the entire APT-MCMC process. It divides all the parallel tempering Markov chains into a series of successive segments, which are referred as stacks in the code, and moniters the acceptance rates obtained at the end of each stack.
   If the acceptance rate of any stack is not within a reasonable range, an additional tuning stage will be scheduled for the corresponding chain.

  \item  \texttt{mpi\_stack.c} $-$ It performs the random walk of each segment chain within a stack.
  It further divides the segment chains in a stack into successive batches, which are even short chains, and accesses the swapping criteria between randomly selected parallel tempering chains after each batch.
  If the criteria for swapping between the chosen parallel chains are met, the current parameter values of the chains will be swapped.

  \item  \texttt{mpi\_batch.c} $-$ It contains the functions that implement the Metropolis-Hastings MCMC algorithm for each batch.

   \item  \texttt{mpi\_tune.c} $-$ This file includes all the functions related to the tuning process, which aims to automatically adjust the sampling step sizes of all parallel tempering chains with bad acceptance rates using a automatic control system.

   \item  \texttt{mpi\_ladder.c} $-$  This file includes a simple algorithm to the tune the ladders of the parallel tempering chains in the beginning tuning stage.

  \item \texttt{data\_loader.c} $-$ This file contains the functions for loading the user data files.

  \item  \texttt{readin.c} $-$ It contains functions for reading the input parameters of Nii-C. These parameters control the overall APT-MCMC process.

  \item \texttt{user\_prior.c} $-$ This is a user-defined file that includes the prior functions of a specific model.

  \item  \texttt{user\_logll.c}  $-$ This is another user-defined file for the likelihood functions of a specific model.

  \item  \texttt{Nii-C.MAN}  $-$ This file describes the important functions in Nii-C to assist the modification of the code if needed.

\end{itemize}

\begin{acknowledgments}
We thank an anonymous referee for providing useful comments that have improved this paper.
S.J. thanks Pablo Pena for helpful discussions. 
This work is supported by National Natural Science Foundation of China (Grant No. 11973094, 12033010, 12103003), Youth Innovation Promotion Association CAS (2020319), the incubation programme for recruited talents (2023GFXK153) and the doctoral start-up funds from Anhui Normal University. 
S.J. is grateful to Yu-Ling Liu and Gui-Mei Wang for their moral support during the development of the Nii-C code.
\end{acknowledgments}

\software{
\texttt{Nii-C} \citep{Jin2024},
\texttt{PTMCMCSampler} \citep{Ellis2019},
\texttt{gnuplot} \citep{Williams2022}, 
\texttt{corner} \citep{Foreman-Mackey2016},
\texttt{Dia}\footnote{\url{http://dia-installer.de}} 
          }

\bibliographystyle{aasjournal}                                                                                                                    
\bibliography{}{}      

\end{document}